\documentclass[9pt,twocolumn,letter]{article}

\usepackage[left=0.5in,right=0.5in,top=0.5in,bottom=1in]{geometry}  
\usepackage{graphics}
\usepackage{graphicx}
\usepackage[colorlinks = true,
            linkcolor = blue,
            urlcolor  = blue,
            citecolor = blue,
            anchorcolor = blue]{hyperref}

\usepackage{mathtools,stmaryrd,soul}
\usepackage{amsmath,amssymb}

\usepackage{authblk}
\usepackage{xcolor}
\usepackage{bm}

\usepackage[utf8]{inputenc}

\newcommand{\beginsupplement}{%
        \setcounter{table}{0}
        \renewcommand{\thetable}{S\arabic{table}}%
        \setcounter{figure}{0}
        \renewcommand{\thefigure}{S\arabic{figure}}%
        \setcounter{section}{0}
        \renewcommand{\thesection}{S\arabic{section}}
     }

\date{}

\begin{document}

\title{Hyperspectral wavefront sensing with a multicore fiber}

\author[1,*]{Baptiste Blochet}
\author[2]{Nathalie Lebas}
\author[1,3,4]{Pascal Berto}
\author[2]{Dimitrios Papadopoulos}
\author[1,4,5,+]{Marc Guillon}

\affil[1]{Saints-Pères Paris Institute for the Neurosciences, CNRS UMR 8003, Université de Paris, 45 rue des Saints-Pères, Paris 75006, France}
\affil[2]{LULI-CNRS, CEA, Sorbonne Universite, Ecole Polytechnique, Institut Polytechnique de Paris, Palaiseau 91128, France}
\affil[3]{Institut de la Vision, Sorbonne Université, CNRS-UMR 7210, Inserm-UMR S968, Paris 75012, France}
\affil[4]{Institut Langevin, ESPCI Paris, Universit\'e PSL, CNRS, Paris 75005, France}
\affil[5]{Institut Universitaire de France, Paris, France}
\affil[*]{baptiste.blochet@u-paris.fr}
\affil[+]{marc.guillon@u-paris.fr}


\maketitle


\paragraph{Abstract.}

\textbf{
Single-shot hyperspectral wavefront sensing is essential for applications like spatio-spectral coupling metrology in high power laser or fast material dispersion imaging. Under broadband illumination, traditional wavefront sensors assume an achromatic wavefront, which makes them unsuitable.
We introduce a hyperspectral wavefront sensing scheme based on the Hartmann wavefront sensing principles, employing  a multicore fiber as a modified Hartmann mask to overcome these limitations. Our system leverages the angular memory effect and spectral decorrelation from the multicore fiber, encoding wavefront gradients into displacements and the spectral information into uncorrelated patterns. This method retains the simplicity, compactness, and single-shot capability of conventional wavefront sensors, with only a slight increase in computational complexity. It also allows a tunable trade-off between spatial and spectral resolution. We demonstrate its efficacy for recording the hyperspectral wavefront cube from single-pulse acquisitions at the Apollon multi-PW laser facility, and for performing multispectral microscopic imaging of dispersive phase objects.
}

\section*{Introduction.}

Wavefront sensing and quantitative phase imaging are widely used for beam metrology, adaptive optics~\cite{hampson2021adaptive} and biomedical imaging~\cite{Park_NP_18}. 
A wavefront sensor typically measures the wavefront of a single monochromatic beam. For broadband beams, it is assumed that the wavefronts are achromatic, otherwise, the wavefront sensor (WFS) can only measure a meaningless spectrally-averaged wavefront. In such cases, hyperspectral wavefront sensing is required. Such spectrally-resolved WFS is typically needed to quantify spatio-spectral coupling in high power laser metrolgy~\cite{jeandet2022survey}, to study light matter interaction in plasma~\cite{tang2022single}, to image material dispersion~\cite{song2022triple} or to measure hemoglobin concentration in living organisms~\cite{park2009spectroscopic}.
In these applications, single-shot measurements are preferable or even required, especially to monitor fast events like in flow cytometry~\cite{lee2019quantitative}, in pump probe experiments~\cite{tang2022single}, in light matter interaction~\cite{wang2020single}, or to characterize single laser pulses~\cite{wang2020single}. Assisted by a chirped pulse, ultrafast events can be acquired~\cite{tang2022single, wang2020single}.
In contrast to hyperspectral phase imaging, multispectral recording only considers a discrete set of wavelengths. Multispectral phase imaging can be achieved by digital holography by encoding spectral information in the k-space~\cite{grace2021single} or even without the need for a reference arm by combining a wavefront sensor with a multispectral camera~\cite{dorrer2018spatio}. Recently, hyperspectral phase maps could be rebuilt experimentally over a continuous spectrum thanks to a phase retrieval algorithm applied to two defocused images acquired with hyperspectral cameras (so-called CASSI)~\cite{tang2022single}. Recent numerical simulations have also paved the way to associating a CASSI imager with a high resolution WFS based on quadriwave lateral shearing interferometry~\cite{howard2023hyperspectral}.
All hyperspectral techniques developed thus far have abandoned the principles of Hartmann-based WFS, consequently forfeiting several of its many advantages that made it highly popular for metrology and quantitative phase imaging~\cite{bon2009quadriwave,baffou2023wavefront,berto2012wide}: simplicity, compactness, single-shot capability, robustness to vibrations, and compatibility with broadband light sources.

Here, we propose a hyperspectral wavefront sensing scheme based on the same principles underlying conventional WFS~\cite{bon2009quadriwave,baffou2023wavefront,berto2012wide}. Specifically, our system leverages both the angular memory effect~\cite{feng1988correlations} and the spectral decorrelation~\cite{goodman2007speckle} occurring at a short distance from the output of a multicore single-mode fiber bundle (MCF). The angular memory effect encodes wavefront gradients in intensity pattern displacements~\cite{berto2017wavefront}, while spectral decorrelation ensures distinct pattern encoding. The proposed instrument thus inherits all the advantages of conventional WFS at the expense of a slightly increased computational post-processing complexity, but does not require any regularization.
We demonstrate the quantitative nature of our approach on well-defined optical systems. Next, we demonstrate the practical usability of our instrument by characterizing the hyperspectral wavefront cube from a single-pulse beam of multi-PW laser line at the Apollon laser facility, captured in a single-shot image acquisition. Finally, since our system relies on Hartmann deflectometry, it enables tuning the trade-off between spatial resolution and phase sensitivity~\cite{berto2017wavefront}. We thus show that our implementation also enables single-shot, multispectral, quantitative phase microscopy with a high spatial resolution.

\section*{Principle}
\paragraph{Image formation.}

\begin{figure*}[h!]
    \centering
    \fbox{\includegraphics[width=1\linewidth]{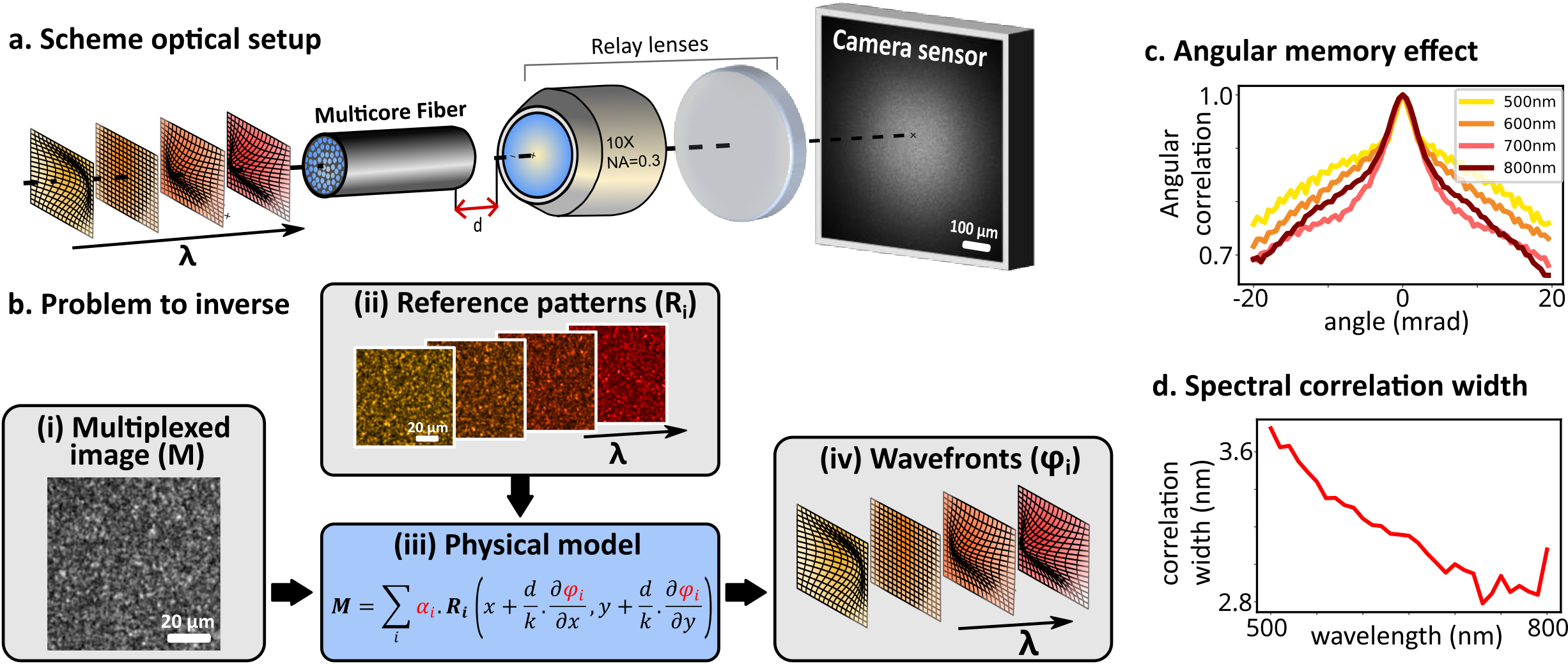}}
    \caption{\textbf{Setup and calibrations. }\textbf{a.} Unknown multispectral wavefronts are imaged onto the input facet of a multicore fiber. A plane at a distance $d$ from the output facet is imaged onto a camera using a $\times10$, $0.3{\rm NA}$ microscope objective and a $f=300~{\rm mm}$ tube lens. \textbf{b.} 
    A multiplexed image (i) consists of an incoherent sum of locally translated reference patterns (ii) according to the model (iii). The set of reference patterns (ii) are sequentially measured under monochromatic and plane wave illumination. The wavefronts (iv) and the intensities are then reconstructed by solving the inverse problem. \textbf{c.} Angular correlation of the wavefront sensor measured at four wavelengths.\textbf{d.} Spectral correlation width of the wavefront sensor as a function of the wavelength.  
}
    \label{fig:shema}
\end{figure*}

The speckle pattern created by a scattering medium exhibiting an angular memory effect~\cite{feng1988correlations} encodes the incident wavefront gradients into local speckle grains displacements at a short distance $d$~\cite{berto2017wavefront, wang2017ultra, sun2022quantitative, wu2024multiplexed, berujon2012two, wu20223d}. For a monochromatic light beam, the relationship between a reference speckle pattern $R(x,y)$ obtained under plane wave illumination and a speckle pattern $M(x,y)$ obtained with unknown beam wavefront and intensity profile is:
\begin{equation}
M(x,y) = \alpha(x,y)\cdot R\left[x+u(x,y), y+v(x,y)\right]
\end{equation}
where $\alpha(x,y)$ is the beam intensity and $u(x,y)$, $v(x,y)$ the components of the vector displacement map induced by the wavefront gradient. The spatial-phase gradient $\nabla_\perp\varphi$ is related to $u(x,y)$ and $v(x,y)$ according to:  
\begin{equation}
d\frac{\nabla_\perp\varphi}{k_0} = u{\bf e_x}+ v{\bf e_y}
\label{eq:grad_displacement}
\end{equation}
Measuring the displacement map $(u,v)$ thus enables the reconstruction of the wavefront $\varphi$ by a numerical integration step~\cite{Asundi_OLE_15,bon2012noniterative,Guillon_APL_21}.

In addition, the spectral decorrelation that arises after propagation through a complex medium encodes spectral information into uncorrelated speckle patterns~\cite{goodman2007speckle}. The possibility to unmix such statistically independent speckle patterns under broadband illumination enables spectroscopy \cite{redding2013all,metzger2017harnessing} and even hyperspectral intensity imaging~\cite{french2017speckle,li2019single}. In these systems, the spectral resolution is determined by the number of interfering spatial modes, which typically scales with the spectral correlation width of the scattering medium, $ \delta\lambda_{medium}$, defining the spectral detuning necessary for speckle decorrelation~\cite{redding2013compact}. 

To achieve hyperspectral wavefront sensing, the Hartmann mask must be carefully chosen. Here we suggest using a MCF which demonstrated both a wide angular memory effect~\cite{porat2016widefield} and a precisely controlled spectral bandwidth~\cite{Rigneault_JOSAB_15,accanto2023flexible}. As illustrated in Fig.~\ref{fig:shema}a and~\ref{fig:shema}b, when a broadband light beam illuminates such a complex medium having angular memory effect, a speckle intensity pattern $M(x,y)$ is created at a distance $d$ from the medium output, which can be imaged on a camera. This speckle is the incoherent sum, over the beam spectrum, of the monochromatic reference speckle patterns $R(x,y,\lambda)$, weighted by the spatially-dependent power spectrum amplitude of the beam intensity $\alpha(x,y,\lambda)$, and distorted according to the incident hyperspectral phase gradients:
\begin{equation}
\label{eq:multiplex}
M(x,y) = \int_\lambda \alpha(x,y,\lambda)R\left[x+u(x,y,\lambda), y+v(x,y,\lambda),\lambda\right] {\rm d}\lambda 
\end{equation}
where $u(x,y,\lambda)$, $v(x,y,\lambda)$ are the displacements induced by the phase gradients $\nabla_\perp\varphi(\lambda)$ according to Eq.~\eqref{eq:grad_displacement} (Fig.~\ref{fig:shema}b). Therefore, because of the spectral diversity of speckles, both the spectro-spatial phases and amplitudes information of the incident field are encoded and multiplexed in the single speckle pattern image $M(x,y)$. 
Since spectral information is spatially encoded in the speckle patterns captured by the camera, increasing spectral resolution requires allocating more camera pixels to each sampling point, which in turn reduces the number of spatial sampling points on the wavefront sensor. Ultimately, this trade-off between the number of spectral and spatial modes is made with the contraint that the total number of modes is limited by the number of camera pixels.

\paragraph{Reconstruction algorithm.}
The inverse problem in Eq.~\eqref{eq:multiplex}, summarized in Fig.~\ref{fig:shema}b, is solved doing the two following assumptions: (i) the displacements $u(x,y,\lambda)$ and $v(x,y,\lambda)$ are sufficiently small, relative to the speckle grain size, to justify a first-order Taylor expansion of Eq.~\eqref{eq:multiplex}. (ii) Both the beam amplitude $\alpha$ and wavefront gradient (\emph{i.e.} $(u,v)$) can be considered as constant within a given neighborhood. This latter assumption is at the basis of the Lucas-Kanade optical flow algorithm~\cite{lucas1981iterative}, a variant of which we use here. Over this neighborhood, Eq.~\eqref{eq:multiplex} can be re-written in the discretized matrix formalism as
%
\begin{equation}
M = \widetilde{R} X + noise
\label{eq:patch}
\end{equation}
where $\widetilde{R} = \left[R_i,\partial_xR_i,\partial_yR_i \right]$ is the matrix composed of all the reference speckles $R_i$ measured at wavelengths $(\lambda_{i})_{i\in [1:N_\lambda]}$ together with their partial derivatives $(\partial_xR_i,\partial_yR_i)$ along $x$ and $y$ coordinates, respectively. $X = [\alpha_i,u_i\alpha_i,v_i\alpha_i]^{T}$ is the sought-for vector containing both the displacements ($u_i,v_i$) and the beam amplitudes $\alpha_i$ at each $\lambda_{i}$ wavelength.  

For a discrete set of wavelength (\emph{i.e.} for multipsectral imaging), the matrix Eq.~\eqref{eq:patch} is solved over a sliding Gaussian window based on a least square minimization by computing the Moore-Penrose pseudo-inverse $X^* = (\widetilde{R}^T\widetilde{R})^{-1}\widetilde{R}M $. When a continuous spectrum is considered (\emph{i.e.} for hyperspectral imaging), Eq.~\eqref{eq:patch} is solved over a set of patches covering the camera surface, using a truncated singular value decomposition (SVD) estimation~\cite{hansen1987truncated} where the low singular values are filtered out with a half-Gaussian function of width $\sigma_{SVD}$~\cite{malone2023diffuserspec}.

The first order Taylor expansion performed to linearize Eq.~\eqref{eq:multiplex} is only valid over displacements of the order of one speckle grain size. To make the resolution robust to larger displacements, we implemented an iterative multi-scale registration algorithm~\cite{weber1995robust}. The principle of the multi-scale approach consists in adjusting the speckle grain size at each iteration by Gaussian-filtering all images ($M$ and $R$), thereby ensuring the first order approximation remains valid at each step by distorting the reference images according to the estimated larger-scaled displacement maps. The filter width $\sigma_{scale}$ is then progressively decreased to measure finer displacement structures. The detailed description of the inversion process, including the definition of the neighborhood and wavelength sampling, can be found in the Methods section.

After solving Eq.~\eqref{eq:patch} over a dense set of patches, a numerical integration of the displacement maps $(u_i,v_i)$ is achieved in the Fourier domain. To avoid artifacts associated with periodic boundary conditions, Fourier integration is achieved after symmetrizing the gradient vector field~\cite{bon2012noniterative}. 

\section*{Results}\label{sec3}

\paragraph{Optical characteristics of our system.} 
The built-up hyperspectral wavefront sensor consists of a MCF (FIGH-100-1500N, Fujikura, Japan), a relay imaging system ($\times10$, $0.3~{\rm NA}$ microscope objective and a $300~{\rm mm}$ tube lens), and a camera (PCO Panda 4.2) that image the output facet of the MCF at a distance $d$.
MCFs exhibit both angular memory effects and spectral decorrelation simultaneously. The spectral decorrelation results from the random dispersion of delays between cores, which is an inherent characteristic of MCF~\cite{accanto2023flexible}. The angular correlation is given by the numerical aperture (NA) of the cores~\cite{porat2016widefield}. In our case, the angular correlation of the fiber bundle exhibit a $\simeq 30\%$ drop for a tilt of $20~{\rm mrad}$ (In Fig.~\ref{fig:shema}c).  Our experiments were conducted within this range, thereby validating this lower bound of the dynamic range for our wavefront sensor implementation. 
The distance $d$ must be tuned so as to control the trade-off between spatial resolution \emph{vs.} wavefront-tilt sensitivity. Increasing $d$ decreases the spatial resolution but improve the sensitivity (according to eq \eqref{eq:grad_displacement}).
Additionally, a lower spatial resolution allows for a larger neighborhood to be considered when solving Eq.~\eqref{eq:patch}, so enabling the use of more pixels to reconstruct additional wavelengths. 
For a distance $d= 88~{\rm mm}$ from the fiber image plane, we measured correlation widths of the order of $3~{\rm nm}$ all over the spectrum $500- 800~{\rm nm}$ (Fig.~\ref{fig:shema}d). Both the angular memory effect and the spectral correlation widths are suitable for hyperspectral wavefront sensing applications.

Here, we demonstrate two implementations of our instrument corresponding to two trade-offs between spatial and spectral resolutions:
\begin{enumerate}
\item One hyperspectral wavefront sensing modality measuring the first few Zernike wavefront modes over $\simeq 20$ spectral channels of a continuous $100~{\rm nm}$-wide spectrum. 

\item One quantitative multispectral and high-resolution wavefront imaging modality ($220\times220~{\rm pixels}$) performed at three discrete wavelengths in a microscopic imaging configuration. 
\end{enumerate}

\paragraph{Quantitative hyperspectral wavefront sensing.} \label{Metro}

\begin{figure*}[h!]
\centering
\fbox{\includegraphics[width=1\linewidth]{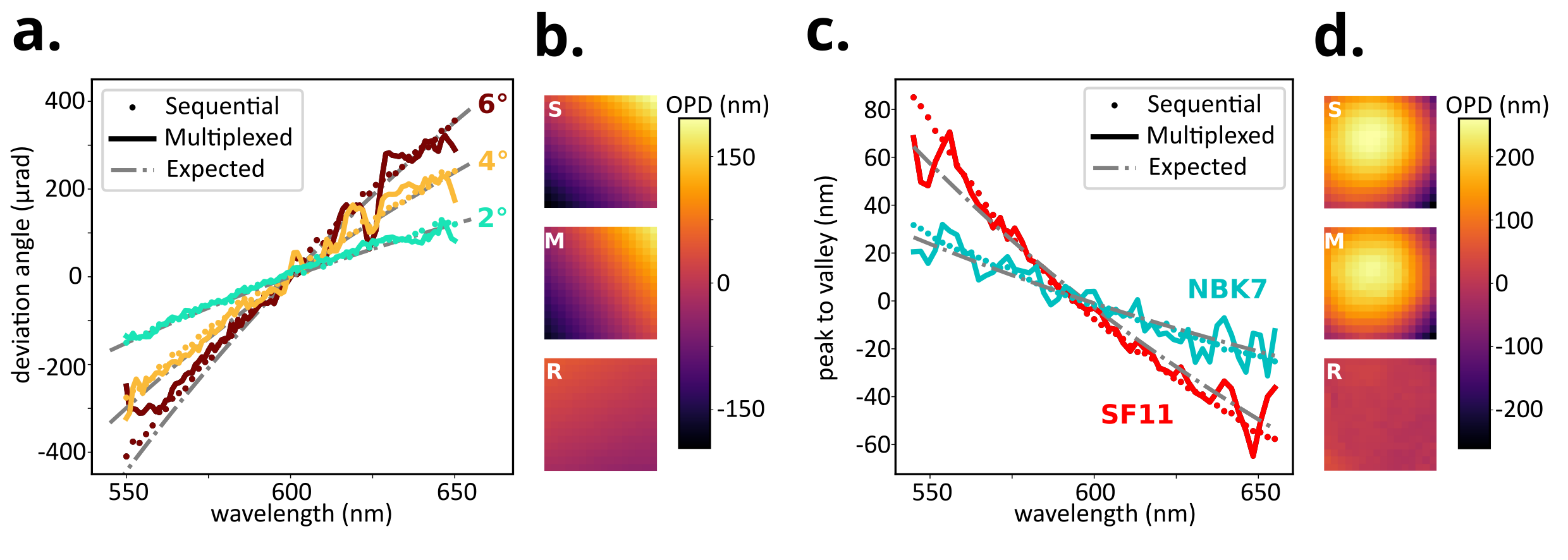}}
\caption{\textbf{Hyperspectral wavefront metrology.}  \textbf{a. }Comparison of the differential deviation angle induced by various prisms ($2^\circ$, $4^\circ$ and $6^\circ$) as a function of the wavelengths, using sequential measurements, single-shot multiplexed measurements and expected values. \textbf{b.} Reconstructed tilted wavefronts at 640nm comparing sequential reconstruction (S), single-shot multiplexed reconstruction (M), and the residual difference between them (R). \textbf{c.} Comparison of the spectral defocus measured for two afocal systems ($f_1=f_2=50~{\rm mm}$) in NBK7 (in cyan) and SF11 (red) as a function of the wavelengths, with sequential measurements, single-shot multiplexed measurements and calculated expected values. \textbf{d.} Reconstructed defocus wavefronts from the SF11 afocal system at $560~{\rm nm}$ comparing sequential measurements (S), single-shot multiplexed (M), and the residual difference between them (R). 
}
\label{fig:metro}
\end{figure*}

To verify the instrument's quantitativity, a first setup was implemented to perform hyperspectral wavefront sensing over a continuous spectrum using simple test optical systems. The test optical systems under consideration were: three wedge prisms made of NBK7 material (with deviation angles of $2^\circ$, $4^\circ$, and $6^\circ$ at $633~{\rm nm}$), and two afocal lens-pair systems ($f = 50~{\rm mm}$) in NBK7 and SF11 materials. A supercontinnum laser coupled with a computer-controled filter box was used as a light source. 
The output plane of the optical system under test was imaged onto the MCF facet using a $1:3$ demagnifying telescope. The distance $d\simeq 90~{\rm mm}$ between the fiber output image plane and the camera 
was finely calibrated over the whole $100~{\rm nm}$-wide spectrum of interest, so taking into account chromatic aberrations of relay optics.  For numerical reconstruction, the patch size at the camera was qualitatively adjusted to match the diffraction pattern of a single fiber after free space propagation, resulting in the reconstruction of phase and intensity maps of resolution $16\times 16$-patches across a $100~{\rm nm}$ spectral range with a spectral resolution of $5~{\rm nm}$.
Next, the wavefronts were reconstructed both from the single broadband multiplexed images and from sequential images acquisition while scanning the spectrum. More details about the experiment and data processing are given in the Methods section.
An illustration of the reconstructed multiplexed and sequential wavefronts is shown in Fig.~\ref{fig:metro}b for the $2^\circ$ prism at $640~{\rm nm}$, while Fig.~\ref{fig:metro}d similarly provides an illustration for the SF11 afocal system at $560~{\rm nm}$. The wavefront accuracy evaluated by computing the root mean square error (RMSE) between the expected and measured wavefronts over the whole spectrum reconstructed from the prisms experiments, is $1.3~{\rm nm}$ (\emph{i.e.} $\lambda/450$ at $600~{\rm nm}$) for the sequential reconstruction and $9.1~{\rm nm}$ (\emph{i.e.} $\lambda/65$ at $600~{\rm nm}$) for the multiplexed reconstruction. 

Finally, we extracted the tip/tilt and defocus components from the wavefronts, for the prisms and afocal systems, respectively, and measured the differential chromatic contribution, taking $600~{\rm nm}$ as an arbitrary reference wavelength. These results are shown in Fig.~\ref{fig:metro}a for the prisms, and in Fig.~\ref{fig:metro}c for the afocal systems. Results obtained from multiplexed measurements exhibit excellent agreement with the ones obtained from sequential data. These experimental results were further compared  with values expected from numerical simulations (based on the geometrical specifications of optics and the tabulated refractive indices of the materials). The theoretical spectral defocus induced by the afocal systems were computed within the frame of the thin lens approximation. 
All measurements show excellent agreement with one another and together with theoretical expectations and demonstrate the quantitative nature of our hyperspectral wavefront sensing approach.

\paragraph{Single pulse reconstruction at the Apollon laser facility.}\label{sec3}

\begin{figure*}[h!]
    \centering
    \fbox{\includegraphics[width=1\linewidth]{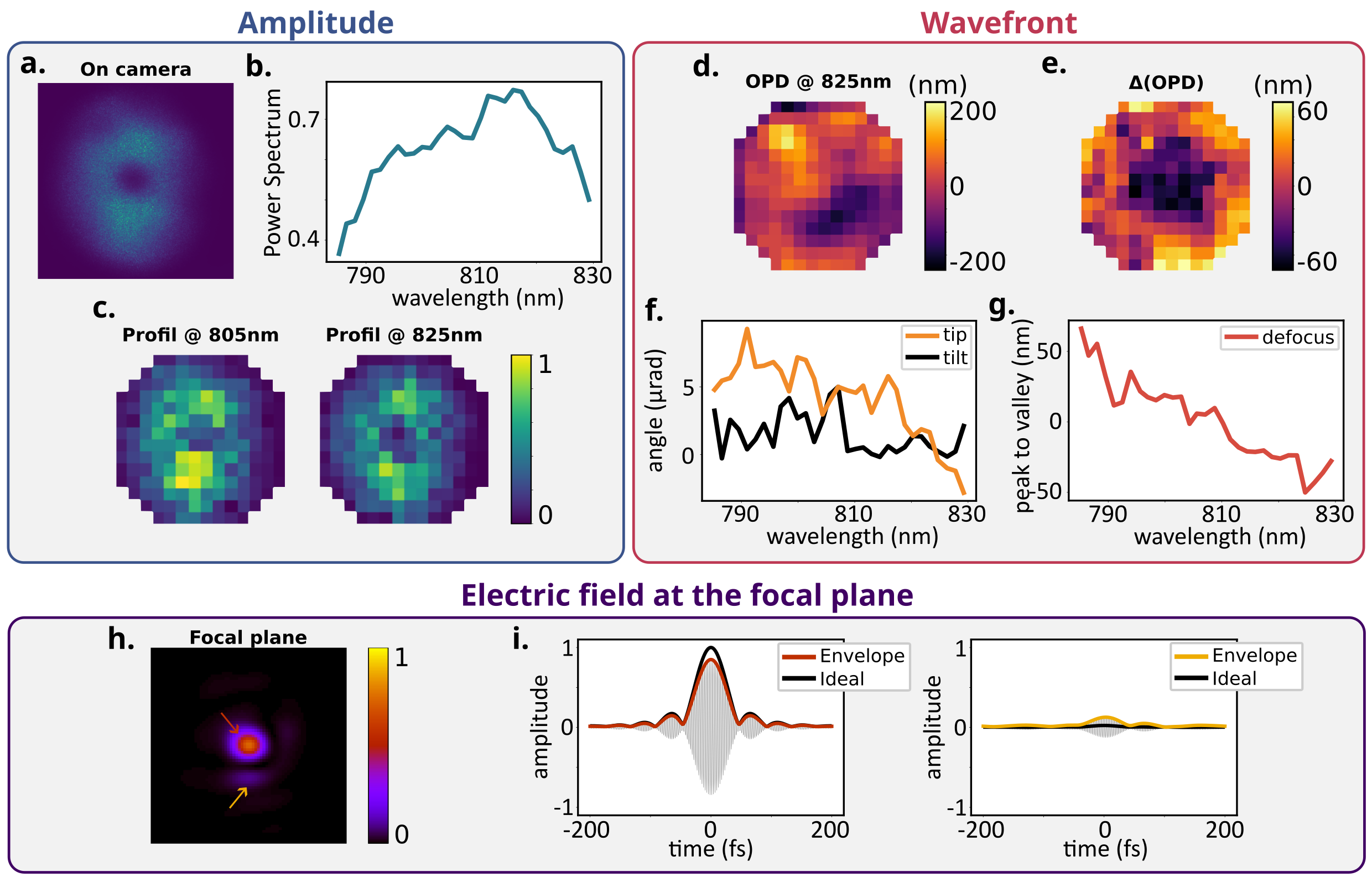}}
    \caption{
    \textbf{Single-pulse reconstruction of a $10~{\rm Hz}$ Apollon pulse. a.} Multiplexed speckle captured by the camera.
    \textbf{b.} Reconstructed power spectrum derived from the multiplexed data.
    \textbf{c. }Intensity profiles reconstructed at $805~{\rm nm}$ and $825~{\rm nm}$.  
    \textbf{d. } Wavefront measured at $825~{\rm nm}$. 
    \textbf{e. } Wavefront difference between $785~{\rm nm}$ and $825~{\rm nm}$.  
    \textbf{f.} Spectral deviation angles across the pulse spectrum. 
    \textbf{g.} Peak-to-valley spectral defocus variation across the pulse spectrum.  
    \textbf{h.} Intensity at the focal plane computed from the measured pupil electric field. The colormap's maximum value is scaled accordint to the peak intensity of an unaberrated focus.
    \textbf{i.} Reconstructed ideal electric fields (dark) and actual electric fields (red-yellow) at the two positions indicated by the arrows in \textbf{h}, assuming a flat spectral phase at the center of the focus. The gray lines represent the real part of the electric field.
    }
    \label{fig:apollon}
\end{figure*}

Next, we installed our single-shot hyperspectral wavefront sensor in the Long Focus Area (LFA) at the Apollon multi-PW laser facility~\cite{papadopoulos2016apollon} (Saint-Aubin, France). 
Wavefront measurements were carried out through the $1~{\rm PW}$ beam-line, using a test laser beam having $50~{\rm mJ}$ pulse energies at $10~{\rm Hz}$ repetition rate (activating 2 amplification stages out of 5).
Since following the very same optical path as in the full-energy configuration (nominal pulse-energy of $15~{\rm J}$ at a rate of $1~{\rm shot/min}$), the laser beam was carrying the main static residual spatiotemporal coupling aberrations of the system. 
The $140~{\rm mm}$ diameter laser beam was demagnified down to a beam diameter of $0.7~{\rm mm}$ at the input facet of the MCF with achromatic telescope relay systems. The hyperspectral wavefront sensor was calibrated \emph{in situ} over the spectral range $650-1000~{\rm nm}$ with a resolution of $1~{\rm nm}$. More details about the system and its calibration can be found in the Methods section.

The multiplexed speckle image is shown in Fig.~\ref{fig:apollon}a. The beam exhibits a central hole resulting from the sampling mirror of the current LFA configuration (see Methods). From this single image, the $\simeq 40~{\rm nm}$-wide spectrum of the laser is retrieved (Fig.~\ref{fig:apollon}b) together with hyperspectral intensity beam profiles (Fig.~\ref{fig:apollon}c) showing very good agreement with the real Apollon beam characteristics. 
The hyperspectral optical path differences (OPD) are also retrieved (see illustration of the OPD in Fig.~\ref{fig:apollon}d for the wavelength $825~{\rm nm}$). Before integration of wavefront gradients, the missing information in the central zero-intensity region was filled in using a median filter.
The measured wavefronts (Fig.~\ref{fig:apollon}d) mostly present low order aberrations. The differential chromatic aberrations between $785~{\rm nm}$ and $825~{\rm nm}$, shown in Fig.~\ref{fig:apollon}e, mostly exhibits chromatic defocus but negligible spectral tilt, so demonstrating the excellent alignement of the Apollon laser system (up to a few ${\rm \mu rad}$ at the WFS plane). The residual spectral tilt and defocus are plotted in Fig.~\ref{fig:apollon}f and Fig.~\ref{fig:apollon}g. 
The chromatic defocus is measured to be about $50~{\rm nm}$ PtV ($\lambda/16$) over the FWHM-bandwidth of the Apollon pulses, corresponding to a $\pm z_R/8$ longitudinal depth of focus, with $z_R$ the Rayleigh length of the focused beam. This value is also in good agreement with the theoretically expected longitudinal chromatism of the Apollon beam transport system employing in its front-end part relay imaging telescopes based on simple singlet fused silica lenses for an aperture up to $18~{\rm mm}$ in diameter~\cite{ranc2022etude}. 

To characterize laser pulses in the temporal domain, measuring the spectral phase is required, at least within a sub-aperture of the near-field beam~\cite{papadopoulos2016apollon}. In the Apollon laser system, the spectral phase is routinely measured and optimized for the central portion of the beam. This enables pulse compression nearly to the Fourier transform limit (FTL). The resulting focus at the focal plane and the electric field time traces at the center of the focus and on one lobe are presented in~\ref{fig:apollon}h and~\ref{fig:apollon}i assuming a flat spectral phase at the center of the focus and preserving the missing hole in the amplitude pupil data. The power spectrum was apodized with a Tukey window with $\alpha=0.35$. The measurements realized in this study are in agreement with previous multi-shot characterization campaigns and indicate negligible peak intensity losses in the order of $15\%$ compared to an ideal focus. 
In contrast to multi-shot acquisition systems, our instrument now enables the characterization of potential non-linear spatio-temporal coupling at the single-pulse level.

\paragraph{Multispectral quantitative phase microscopy.}

\begin{figure*}[h!]
    \centering
    \fbox{\includegraphics[width=1\linewidth]{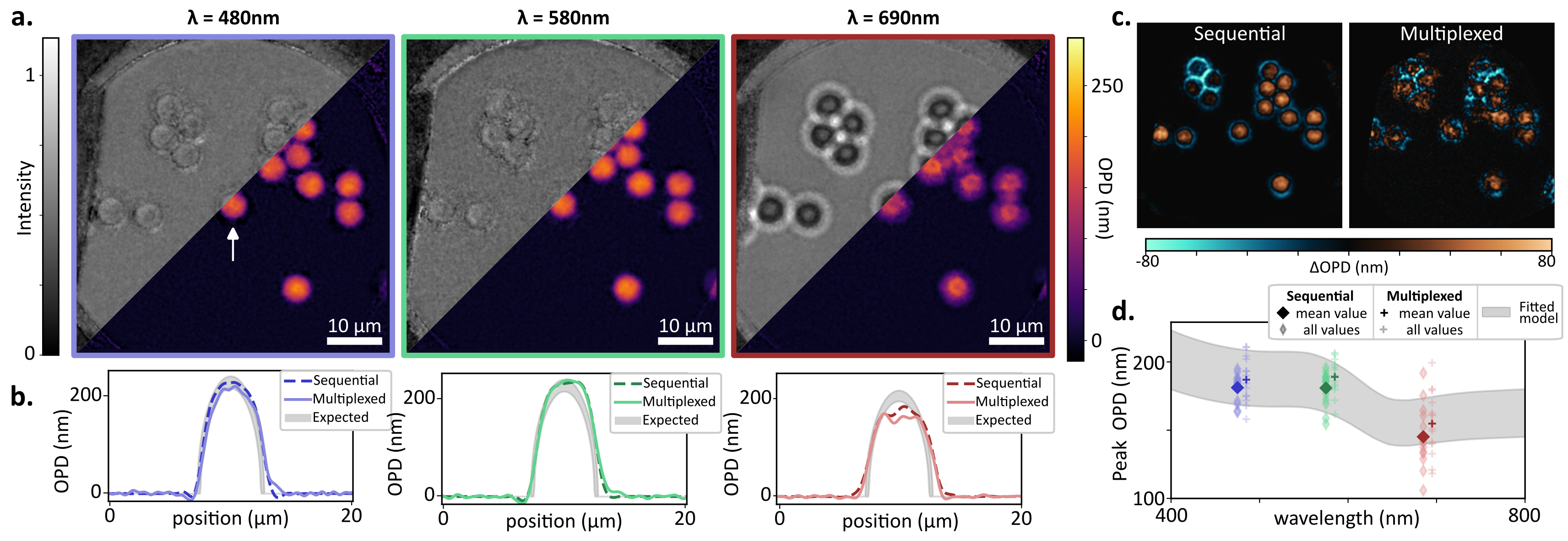}}
    \caption{\textbf{Single-shot high resolution multispectral reconstruction of 5.2um PMMA beads at $480/20~{\rm nm}$ (blue), $580/10~{\rm nm}$ (green) and $690/10~{\rm nm}$ (red).}  \textbf{a.} Intensity and OPD measurements captured at the three wavelengths in single-shot. Scale bar $= 10~{\rm \mu m}$. \textbf{b.} OPD cross-sections of the bead located at the center of the field of view (white arrow in \textbf{a}), after numerical refocusing from sequential (dashed line) and multiplexed (solid line) reconstructions. For comparison, theoretical expectation is shown as the grey strip. \textbf{c.} Differential OPD between the blue ($480~{\rm nm}$) and red ($690~{\rm nm}$) channels. \textbf{d.} Measured peak OPDs, fitted by a dispersion model considering the contribution of both PMMA and the solution (made of water, sucrose, and FCF chromophore).
}
\label{fig:HR}
\end{figure*}

Single-shot multispectral imaging is also of interest for microscopy applications. We thus implemented our system in high-spatial resolution working regime by reducing the distance $d$ down to $\approx 6~{\rm mm}$ between the image of the fiber-bundle output and the camera. The distances $d$ were accurately measured over the full field of view (Supp.~Fig.~\ref{fig:zmaps}). The sample was illuminated in transmission mode with a custom-built source based on three dielectric bandpass filters ($480/20~{\rm nm}$, $580/10~{\rm nm}$ and $690/10~{\rm nm}$), and imaged onto the PCF using a $\times20$ magnification and $0.4~{\rm NA}$ microscope objective. The PSF sampling factor by the MCF is dictated by the $4.5~{\rm \mu m}$ mean nearest-neighbor distance between fiber cores and varies in the range between $2.7$ (at $480~{\rm nm}$) and $3.8$ (at $690~{\rm nm}$). Eq.~\eqref{eq:patch} was solved using a Moore-Penrose pseudo-inversion over a sliding Gaussian window with standard deviation $\sigma_{Gauss}=4~{\rm pixels}$. 
The sample was prepared, consisting in $5.2~{\rm \mu m}$ PMMA beads immersed in an aqueous solution. Sucrose was added to the solution to reduce the refractive index mismatch between water and PMMA. A chromophore (BlueBrillant FCF) was also added in order to introduce a controled amount of spectral dispersion~\cite{sai2020designing}. All details regarding the optical system, the sample preparation and the algorithm can be found in the Methods section. 
Then, multiplexed and sequential multispectral wavefront imaging of the sample was carried out. 

The single-shot multispectral intensities and OPDs are shown in Fig.~\ref{fig:HR}a. Noteworthy, the imaging system between the sample and the WFS exhibits a slight chromatic aberrations since the red channel ($690~{\rm nm}$) appears unfocused. Based on both the measured intensity and OPDs, numerical refocusing is performed (Supp.~Fig.~\ref{fig:postproc}). Comparison of OPD profiles obtained by sequential and multiplexed measurements can then be done and confronted to theoretical expectations based on the solution composition and the specified geometric characteristics of the beads (Fig.~\ref{fig:HR}b). The only missing parameters, the exact sucrose and chromophore concentrations, were fitted from experimental data. The dispersion of the sample clearly appears in the images shown in Fig.~\ref{fig:HR}c where the difference in OPD at $480~{\rm nm}$ and $690~{\rm nm}$ is displayed. Finally, the peak OPDs are plotted in Fig.~\ref{fig:HR}d both from sequential and multiplexed data, demonstrating very good agreement between experimental results and the spectral dispersion expected from theoretical modeling. The scattering observed in the measurements of refractive indices is due to the dispersion in the sizes of the micro-beads (Supp.~Fig.~\ref{fig:peak_OPD}).

\section*{Discussion and conclusion}\label{sec12}

In summary, we have introduced a hyperspectral wavefront sensor with an accuracy of approximately $\lambda/60$ and a tunable spatio-spectral resolution. The sensitivity, defined as the minimum measurable angle, is proportional to the minimal detectable displacement using the implemented optical flow algorithm, which is about $1/200^{\rm th}$ pixel. For the two proposed implementations, the sensitivity is $7.5~{\rm \mu m}$ and  $0.35~{\rm \mu m}$ for the high spatial resolution and high spectral resolution experiments, respectively.

The maximum achievable spectral resolution is determined by the spectral correlation width of the MCF ($\delta\lambda_{medium}\approx 3~{\rm nm}$), and can therefore be changed by tuning the fiber length. Practically, the effective spectral resolution is controlled by the width of the half-Gaussian filter $\sigma_{SVD}$ used to truncate the singular values. The number of recovered spectral channels is then given by $N_{\lambda} = 3\,\sigma_{laser}\diagup\sigma_{SVD}$, where  $\sigma_{laser}$ is the beam bandwidth, and the factor $3$ accounts for the three independent information recovered: the amplitude and the phase gradient components. Based on our experimental results, we assess that the spectral resolution of the hyperspectral wavefront sensor instrument can be qualitatively estimated as $2\sigma_{SVD}$.

The spatial resolution is limited by the free-space propagation distance $d$ between the output facet of the MCF and the camera sensor. At short distances, in the high-spatial resolution microscopy implementation, the spatial resolution is limited by the mean inter-core distance equal to $4.5~{\rm \mu m}$. In the high-sensitivity mode used for laser and optics metrology where the distance $d$ is large, the spatial resolution is $66~{\rm \mu m}$ at the output facet of the MCF. For quantitative phase microscopy, the point spread function of the incoming beam (after magnification) should be properly sampled by the fiber cores.
In our current setup, the speckle grains measured on the camera are oversampled ($12\times12~{\rm pixels}$), suggesting that a larger field of view could be achieved with a bundle containing more fibers. Moreover, the small size of the current MCF ($0.7~{\rm mm}$ in diameter) required a relay imaging system, which introduces chromatic aberrations that must be taken into account through an accurate calibration of the system. 

The post-processing time to unmix the spectral components and compute the intensity and wavefront images was $5~{\rm min.}$ in the laser and optics metrology experiments, and $50~{\rm sec.}$ per scale for the high spatial resolution implementation, with the matrix inversion being the most time-consuming step. In this latter configuration involving large dynamics of the wavefront gradient, the multiscale processing algorithm, requiring multiple matrix inversions, is required for phase imaging to be quantitative. However, for qualitative real-time monitoring, image processing can be achieved on a single scale wherein the matrix inversion step, only requiring the calibration patterns, can be performed once for all and in advance. Optimizing the reconstruction speed, combined with the use of a full-field spatial light modulator~\cite{mounaix2020time,chen2022synthesizing,cruz2022synthesis}, could enable real-time adaptive optics to correct chromatic aberrations and spatio-temporal coupling.

As a perspective, the hyperspectral wavefront sensor presented here appears as perfectly suited for characterizing the spatio-temporal coupling effects of single pulses in ultrashort laser installations like Apollon. Single-pulse metrology is especially critical in this case since Apollon operate at a rate of only 1 shot per minute at full power (both for the 1 PW and 10 PW beam-lines). Our instrument now enables the investigation of potential nonlinear chromatic effects~\cite{Papadopoulos_2022}, and a more precise and comprehensive estimation of the on-the-target focused intensity. Additionally, the high spatial resolution system could be combined with tomographic microscopes~\cite{sung2023hyperspectral} to extract not only the optical path delay dispersion but also the refractive index dispersion of a sample.




\section*{Acknowledgments}\label{sec13}
The authors acknoweldge the support from the team responsible for the Apollon laser facility. They also acknowledge Dan Oron, Ori Katz and Fabrice Harms for stimulating discussions and Christophe Tourain for helping with the preparation of MCFs.
 
\section*{Methods}
\label{sec:methods}


\paragraph{Optical design of the wavefront sensor.} 
The output facet of a $1~{\rm cm}$-long multicore few-mode fiber (FIGH-100-1500N, Fujikura, Japan) is imaged at a distance $d$ from a camera (PCO Panda 4.2) with a microscope objective ($\times10$, $NA = 0.3$) on a translation stage and a tube lens ($f=300~{\rm mm}$ and $f=200~{\rm mm}$ for the hyperspectral system and multispectral system respectively), as illustrated in Fig.~\ref{fig:shema}a. The fraction of transmitted light energy through the MCF is $95\%$. 


\paragraph{Light sources.} 
For hyperspectral WFS, the optical system is illuminated using narrow linewidth collimated beams generated by broadband laser sources and filtered using computer-driven filter boxes. For the prism/afocal experiment
, the supercontinuum source (Leukos, Electro Vis) and the filter box (Leukos, Bepop) provided a minimum bandwidth of $\simeq 5~{\rm nm}$ in the spectral range between $500~{\rm nm}$ and $800~{\rm nm}$. 
At the Apollon laser, a Ti:Sa laser  (Rainbow, Femtolasers Produktions GmbH, Austria) covering the spectral range $650-1000~{\rm nm}$ was used for calibration, in association  with a custom-built monochromator. The monochromator consisted of a pair of blazed gratings, an afocal telescope and a slit mounted on a motorized translation stage. After the monochromator, the beam had a linewidth of $\simeq 2~{\rm nm}$ over the spectral range $730-860~{\rm nm}$. 
The calibration of the monochromators was completed, and their performance was evaluated using a spectrometer (Ocean, ST VIS-25). 
At the output of the monochromators the beams were spatially filtered using single mode fibers and re-collimated before being sent on the input facet of the MCF or the test optics. 
The reference speckle patterns $R_i$ were then sequentially recorded over the given spectral ranges by steps of $1~{\rm nm}$. Broadband illumination could be obtained by tuning the linewidth up to $100~{\rm nm}$ for the commerical filter box, and simply by removing the slit from the custom-built monochromator.

For the multispectral wavefront imaging experiment, a white halogen lamp, spatially filtered by a pinhole and collimated, was split into three optical paths. The three illuminating wavelengths were obtained by inserting bandpass filters ($480/20~{\rm nm}$, $580/10~{\rm nm}$ and $690/10~{\rm nm}$), placed in each path before recombination using beam-splitters.

\paragraph{Spectral calibration of the hyperspectral WFS.}
Multicore single-mode fibers (MCFs) exhibit spectral decorrelation due to the inherent random dispersion of delays between cores~\cite{accanto2023flexible}. 
The spectral correlation width of the MCF was measured over the full spectrum by measuring the speckle contrast $C$ as a function of the central wavelength of the $5~{\rm nm}$-wide laser line. 
The contrast is directly related to the number of independent speckles $N_\lambda$ present in the image as: $N_\lambda = 1/C^{2}$~\cite{goodman2007speckle,curry2011direct}. The speckle contrast decreases as the number of independent spectral speckles increases. We thus conventionally define the spectral correlation width of the MCF as $ \delta\lambda_{medium} = \sigma_{laser} / N_{\lambda}$ where $\sigma_{laser}$ is the linewidth of the laser used. As shown in Fig.~\ref{fig:shema}d, we measured correlation widths of the order of $3~{\rm nm}$ over the full spectrum in between $500~{\rm nm}$ and $800~{\rm nm}$. The spectral width of the MCF ultimately determines the maximum spectral resolution of our hyperspectral WFS. 

\paragraph{Angular memory effect \& calibration of the distance $d$.}
Multicore single-mode fibers (MCFs) exhibit an angular memory effect.
The angular correlation, limited by the numerical aperture (NA) of the cores~\cite{porat2016widefield} was measured experimentally. The tilt angle of a collimated beam was scanned at the input facet of the MCF. The tilt angle was controlled by translating the output of a monomode fiber in the focal plane of a fixed collimating lens ($f=30~{\rm mm}$) and the output of the lens was then conjugated with the input facet of the MCF using a $G_{in} = 1:6$ magnification afocal telescope. The output speckle measured at a distance $d$ on the camera were then digitally cross-correlated to measure the angular memory effect. 
The results shown in Fig.~\ref{fig:shema}c exhibit a $\simeq 30\%$ drop in correlation after a $20~{\rm mrad}$ tilt, over the full spectrum of interest. This limited angular memory effect reduces our ability to recognize speckles and reconstruct wavefronts, thereby restricting the number of accessible spectro-spatial modes. 

The described system also allowed an accurate measurement of the distance $d$ between the output facet image and the camera, accounting for potential chromatic defocus in the relay imaging system. The distance $d$ was measured using Eq.~\ref{eq:grad_displacement}, relying on tilt-angle conservation between the input and the output facets of the MCF, and within the range of the angular memory effect. 
Considering the magnification $G_{in}$ and $G_{out}$ of the relay imaging system before and after the MCF, respectively, the distance $d$ is measured according to:
\begin{equation}
d = f\frac{G_{in}}{G_{out}}\frac{\left<\delta u_{cam}\right>}{\delta u_{source}}
\label{eq:d_calib}
\end{equation}
where $\left<\delta u_{cam}\right>$ is the mean displacement measured over the camera and $\delta u_{source}$ the known displacement amount of the monomode fiber used as a point source. 
For the hyperspectral experiments, an average distance $d$ across the field of view was measured. In the multispectral experiments, to achieve metrology-grade phase imaging, the distance $d$ was estimated individually for each camera pixel. The resulting $d$-maps obtained for each wavelength were fitted using low-order ($N=3$) 2D polynomials, as shown in Fig.~\ref{fig:zmaps}.

\paragraph{Description of the hyperspectral wavefront sensing experiments.}

The metrological measurements on the prisms and afocal systems where carried out by imaging the test optics with afocal imaging telescope with de-magnification $1:3$, onto the MCF. 
Special care was taken to align the test beams with the calibrating beams at the entrance of the MCF. The imaged field of view corresponded to an area of $2.4\times 2.4~{\rm mm}^2$ of the test optics.  
The distance $d$ between the output facet image plane and the camera was accurately measured, and was observed to vary from $88~{\rm mm}$ to $97~{\rm mm}$ over the spectrum of interest ($550~{\rm nm}$ to $650~{\rm nm}$). This relatively large mean distance was intentionally chosen to enhance sensitivity to small wavefront distortions, such as those produced by chromatic optical elements that simulate spatio-temporal coupling in typical ultrashort laser facility systems. 

\paragraph{Description of the Apollon laser beam metrology.}
For hyperspectral WFS acquisitions, the $140~{\rm mm}$ diameter Apollon laser beam was sampled after the focal plane of the end-chain focusing optic ($F=6~{\rm m}$) and collimated to $7~{\rm mm}$ diameter beam using afocal telescopes within the interaction chamber of the LFA. Relay imaging systems, composed of achromatic lenses both in vacuum and in air, were employed to transfer the image plane (defined at the entrance of the experimental chamber) to the input of the hyperspectral WFS. The overall demagnification factor of the imaging system was $1:200$, reducing the beam diameter to $0.7~{\rm mm}$ at the input facet of the MCF. The corresponding field of view at the fiber bundle facet is $0.8\times 0.8~{\rm mm}^2$. 

\paragraph{Description of the multispectral wavefront microscope.}

Poly(methyl methacrylate) beads of $5.2~{\rm \mu m}$ (PMMA-R-5.2, microparticles GmbH) were deposited on glass microscope slides. The dry beads were then covered with a solution consisting in blue brilliant absorber FCF in water (E133 in water, Vahine, Avignon, France) mixed with sucrose ($2~{\rm g/ml}$) and then sealed with a cover glass. A description of the model used to describe our sample refractive index can be found in Supplementary Materials. 
The beads sample was imaged onto the input plane of the wavefront sensor by a microscope imaging system, consisting of a $\times20$- $0.4~{\rm NA}$ microscope objective and a $180~{\rm mm}$ tube lens. The resulting field of view was $120\times 120~{\rm \mu m}^2$. 
The distances $d$ between the output facet image and the camera were accurately measured over the full field of view and varied in the range between $4~{\rm mm}$ and $8~{\rm mm}$ (see Fig.~\ref{fig:zmaps}) after taking into account the magnification after the MCF.

\paragraph{Data processing.}

For multispectral reconstruction involving a limited number of wavelengths with spectral shifts exceeding several $ \delta\lambda_{medium}$, Eq.~\eqref{eq:patch} can be addressed using the Moore-Penrose pseudoinverse: $X^* = (\widetilde{R}^T\widetilde{R})^{-1}\widetilde{R}M $ applied over a neighborhood defined by a Gaussian sliding windows. The full width at half maximum of the gaussian sliding window is selected to ensure that the condition number of the matrix $\widetilde{R}$ remains below 10. Typically, for reconstruction using three discrete wavelengths, this size corresponds to the average distance between two speckle grains observed at the camera plane.

For hyperspectral wavefront reconstruction over a continuous spectrum, the references are captured for a discrete set of wavelengths having a step smaller than the spectral correlation width of the MCF. We found that good performances were obtained for a spectral step  equals to $\delta\lambda_{medium}/3$. We solved Eq.~\eqref{eq:patch} for a dense set of square patches covering the full camera surface.  To solve Eq.~\eqref{eq:patch} for a given patch, the corresponding matrix $\widetilde{R}$  is inverted through a truncated singular value decomposition (SVD) \cite{hansen1987truncated}. The truncation aims at canceling the contribution of low singular values that dominates after inversion. They are thus filtered out by multiplying inverted singular values by a (half-)Gaussian function of width $\sigma_{SVD}$~\cite{malone2023diffuserspec}. In practice, the value of $\sigma_{SVD}$ is set empirically in order to reduce the oscillation-like artefacts appearing during the matrix inversion~\cite{shinn2023phantom}. Furthermore, to speed up the processing time, a sparse SVD routine is used. 
The number of singular values conserved after SVD filtering (the $\sigma_{SVD}$ value) will, \emph{in fine}, drive the maximum spectral resolution of the system. Three singular values are needed per wavelength-band $\lambda_i\pm\delta\lambda_i/2$ to reconstruct all the information: one singular value for the amplitude $\alpha_i$ and two for the displacements $(u_i,v_i)$. With our optical instrument, we used $\sigma_{SVD}=30$ for a full spectral range spanning over $100~{\rm nm}$. 

Both for multispectral and hyperspectral phase imaging, the first order Taylor expansion performed to linearize Eq.~\eqref{eq:multiplex} is only valid over displacements of the order of one speckle grain size. For larger displacements, we implemented an iterative multi-scale version~\cite{weber1995robust}. The principle of the multi-scale approach consists in adjusting the speckle grain size at each iteration by Gaussian-filtering all images ($M$ and $R$). At each iteration, the reference images are distorted according to the estimated large-scaled displacement maps. The filter width $\sigma_{scale}$ is then progressively decreased to measure finer displacement structures.

For the images acquired with the Apollon laser facility, an initial registration step was achieved by patches by digital image cross-correlation between the acquired multiplexed speckle and a synthetic broadbrand reference (\emph{i.e.} the incoherent sum of the measured references) to compensate a displacement offset between the reference speckles and the multiplexed ones. This offset was due to a residual non co-linearity error between the Apollon laser beam and the calibrating beam.

A summary of the parameters used for the reconstructions, along with the resulting imaging characteristics, is provided in Supp.~Table~\ref{table:algo}.

\paragraph{Image processing for multispectral microscopy.}

Additional post-processing was necessary to evaluate the dispersion of the PMMA bead sample in a BlueBrillant solution. Due to chromatic aberrations in the microscope imaging system, all three channels could not be simultaneously focused. Furthermore, air flow fluctuations in the experimental room introduced low-frequency background variations in the reconstructed optical path delays, with amplitudes comparable to the spectral dispersion. 
Defocus was corrected using the angular spectrum method within the Fresnel approximation. Assuming that the objects are transparent, the optimal numerical propagation distance was determined by minimizing the standard deviation of intensity images. The slowly-varying wavefront background was estimated by segmenting the image. At the bead positions, an interpolation was performed. This measured background was then filtered using a Gaussian low-pass filter with a full-width half-maximum of $5$~pixels, and then subtracted from the optical path delay images. The complete reconstruction workflow is illustrated in Fig.~\ref{fig:postproc}. Finally the peak values of OPDs were measured at the center of each bead by averaging over a disk with a radius of $10$~pixels ($=0.6~{\rm \mu m}$).


\bibliographystyle{ieeetr}

\begin{thebibliography}{10}

\bibitem{hampson2021adaptive}
K.~M. Hampson, R.~Turcotte, D.~T. Miller, K.~Kurokawa, J.~R. Males, N.~Ji, and
  M.~J. Booth, ``Adaptive optics for high-resolution imaging,'' {\em Nature
  Reviews Methods Primers}, vol.~1, no.~1, p.~68, 2021.

\bibitem{Park_NP_18}
Y.~Park, C.~Depeursinge, and G.~Popescu, ``Quantitative phase imaging in
  biomedicine,'' {\em Nature Photonics}, vol.~12, pp.~578--589, Oct 2018.

\bibitem{jeandet2022survey}
A.~Jeandet, S.~W. Jolly, A.~Borot, B.~Bussi{\`e}re, P.~Dumont, J.~Gautier,
  O.~Gobert, J.-P. Goddet, A.~Gonsalves, A.~Irman, {\em et~al.}, ``Survey of
  spatio-temporal couplings throughout high-power ultrashort lasers,'' {\em
  Optics express}, vol.~30, no.~3, pp.~3262--3288, 2022.

\bibitem{tang2022single}
H.~Tang, T.~Men, X.~Liu, Y.~Hu, J.~Su, Y.~Zuo, P.~Li, J.~Liang, M.~C. Downer,
  and Z.~Li, ``Single-shot compressed optical field topography,'' {\em Light:
  Science \& Applications}, vol.~11, no.~1, p.~244, 2022.

\bibitem{song2022triple}
J.~Song, J.~Min, X.~Yuan, Y.~Xue, C.~Bai, and B.~Yao, ``Triple-wavelength
  quantitative phase imaging with refractive index measurement,'' {\em Optics
  and Lasers in Engineering}, vol.~156, p.~107110, 2022.

\bibitem{park2009spectroscopic}
Y.~Park, T.~Yamauchi, W.~Choi, R.~Dasari, and M.~S. Feld, ``Spectroscopic phase
  microscopy for quantifying hemoglobin concentrations in intact red blood
  cells,'' {\em Optics letters}, vol.~34, no.~23, pp.~3668--3670, 2009.

\bibitem{lee2019quantitative}
K.~C. Lee, M.~Wang, K.~S. Cheah, G.~C. Chan, H.~K. So, K.~K. Wong, and K.~K.
  Tsia, ``Quantitative phase imaging flow cytometry for ultra-large-scale
  single-cell biophysical phenotyping,'' {\em Cytometry Part A}, vol.~95,
  no.~5, pp.~510--520, 2019.

\bibitem{wang2020single}
P.~Wang, J.~Liang, and L.~V. Wang, ``Single-shot ultrafast imaging attaining 70
  trillion frames per second,'' {\em Nature communications}, vol.~11, no.~1,
  p.~2091, 2020.

\bibitem{grace2021single}
E.~Grace, T.~Ma, Z.~Guang, R.~Jafari, J.~Park, J.~Clark, G.~Kemp, J.~Moody,
  M.~Rhodes, Y.~Ping, {\em et~al.}, ``Single-shot complete spatiotemporal
  measurement of terawatt laser pulses,'' {\em Journal of Optics}, vol.~23,
  no.~7, p.~075505, 2021.

\bibitem{dorrer2018spatio}
C.~Dorrer and S.-W. Bahk, ``Spatio-spectral characterization of broadband
  fields using multispectral imaging,'' {\em Optics express}, vol.~26, no.~25,
  pp.~33387--33399, 2018.

\bibitem{howard2023hyperspectral}
S.~Howard, J.~Esslinger, R.~H. Wang, P.~Norreys, and A.~D{\"o}pp,
  ``Hyperspectral compressive wavefront sensing,'' {\em High Power Laser
  Science and Engineering}, vol.~11, p.~e32, 2023.

\bibitem{bon2009quadriwave}
P.~Bon, G.~Maucort, B.~Wattellier, and S.~Monneret, ``Quadriwave lateral
  shearing interferometry for quantitative phase microscopy of living cells,''
  {\em Optics express}, vol.~17, no.~15, pp.~13080--13094, 2009.

\bibitem{baffou2023wavefront}
G.~Baffou, ``Wavefront microscopy using quadriwave lateral shearing
  interferometry: From bioimaging to nanophotonics,'' {\em ACS photonics},
  vol.~10, no.~2, pp.~322--339, 2023.

\bibitem{berto2012wide}
P.~Berto, D.~Gachet, P.~Bon, S.~Monneret, and H.~Rigneault, ``Wide-field
  vibrational phase imaging,'' {\em Physical Review Letters}, vol.~109, no.~9,
  p.~093902, 2012.

\bibitem{feng1988correlations}
S.~Feng, C.~Kane, P.~A. Lee, and A.~D. Stone, ``Correlations and fluctuations
  of coherent wave transmission through disordered media,'' {\em Physical
  review letters}, vol.~61, no.~7, p.~834, 1988.

\bibitem{goodman2007speckle}
J.~W. Goodman, {\em Speckle phenomena in optics: theory and applications}.
\newblock Roberts and Company Publishers, 2007.

\bibitem{berto2017wavefront}
P.~Berto, H.~Rigneault, and M.~Guillon, ``Wavefront sensing with a thin
  diffuser,'' {\em Optics Letters}, vol.~42, no.~24, pp.~5117--5120, 2017.

\bibitem{wang2017ultra}
C.~Wang, X.~Dun, Q.~Fu, and W.~Heidrich, ``Ultra-high resolution coded
  wavefront sensor,'' {\em Optics express}, vol.~25, no.~12, pp.~13736--13746,
  2017.

\bibitem{sun2022quantitative}
J.~Sun, J.~Wu, S.~Wu, R.~Goswami, S.~Girardo, L.~Cao, J.~Guck, N.~Koukourakis,
  and J.~W. Czarske, ``Quantitative phase imaging through an ultra-thin
  lensless fiber endoscope,'' {\em Light: Science \& Applications}, vol.~11,
  no.~1, p.~204, 2022.

\bibitem{wu2024multiplexed}
T.~Wu, M.~Guillon, G.~Tessier, and P.~Berto, ``Multiplexed wavefront sensing
  with a thin diffuser,'' {\em Optica}, vol.~11, no.~2, pp.~297--304, 2024.

\bibitem{berujon2012two}
S.~B{\'e}rujon, E.~Ziegler, R.~Cerbino, and L.~Peverini, ``Two-dimensional
  x-ray beam phase sensing,'' {\em Physical review letters}, vol.~108, no.~15,
  p.~158102, 2012.

\bibitem{wu20223d}
T.~Wu, M.~Guillon, C.~Gentner, H.~Rigneault, G.~Tessier, P.~Bon, and P.~Berto,
  ``3d nanoparticle superlocalization with a thin diffuser,'' {\em Optics
  Letters}, vol.~47, no.~12, pp.~3079--3082, 2022.

\bibitem{Asundi_OLE_15}
L.~Huang, M.~Idir, C.~Zuo, K.~Kaznatcheev, L.~Zhou, and A.~Asundi, ``Comparison
  of two-dimensional integration methods for shape reconstruction from gradient
  data,'' {\em Optics and Lasers in Engineering}, vol.~64, pp.~1--11, 2015.

\bibitem{bon2012noniterative}
P.~Bon, S.~Monneret, and B.~Wattellier, ``Noniterative boundary-artifact-free
  wavefront reconstruction from its derivatives,'' {\em Applied optics},
  vol.~51, no.~23, pp.~5698--5704, 2012.

\bibitem{Guillon_APL_21}
T.~Wu, P.~Berto, and M.~Guillon, ``{Reference-less complex wavefields
  characterization with a high-resolution wavefront sensor},'' {\em Applied
  Physics Letters}, vol.~118, p.~251102, 06 2021.

\bibitem{redding2013all}
B.~Redding, S.~M. Popoff, and H.~Cao, ``All-fiber spectrometer based on speckle
  pattern reconstruction,'' {\em Optics express}, vol.~21, no.~5,
  pp.~6584--6600, 2013.

\bibitem{metzger2017harnessing}
N.~K. Metzger, R.~Spesyvtsev, G.~D. Bruce, B.~Miller, G.~T. Maker, G.~Malcolm,
  M.~Mazilu, and K.~Dholakia, ``Harnessing speckle for a sub-femtometre
  resolved broadband wavemeter and laser stabilization,'' {\em Nature
  communications}, vol.~8, no.~1, p.~15610, 2017.

\bibitem{french2017speckle}
R.~French, S.~Gigan, and O.~L. Muskens, ``Speckle-based hyperspectral imaging
  combining multiple scattering and compressive sensing in nanowire mats,''
  {\em Optics letters}, vol.~42, no.~9, pp.~1820--1823, 2017.

\bibitem{li2019single}
X.~Li, J.~A. Greenberg, and M.~E. Gehm, ``Single-shot multispectral imaging
  through a thin scatterer,'' {\em Optica}, vol.~6, no.~7, pp.~864--871, 2019.

\bibitem{redding2013compact}
B.~Redding, S.~F. Liew, R.~Sarma, and H.~Cao, ``Compact spectrometer based on a
  disordered photonic chip,'' {\em Nature Photonics}, vol.~7, no.~9,
  pp.~746--751, 2013.

\bibitem{porat2016widefield}
A.~Porat, E.~R. Andresen, H.~Rigneault, D.~Oron, S.~Gigan, and O.~Katz,
  ``Widefield lensless imaging through a fiber bundle via speckle
  correlations,'' {\em Optics express}, vol.~24, no.~15, pp.~16835--16855,
  2016.

\bibitem{Rigneault_JOSAB_15}
E.~R. Andresen, S.~Sivankutty, G.~Bouwmans, L.~Gallais, S.~Monneret, and
  H.~Rigneault, ``Measurement and compensation of residual group delay in a
  multi-core fiber for lensless endoscopy,'' {\em J. Opt. Soc. Am. B}, vol.~32,
  pp.~1221--1228, Jun 2015.

\bibitem{accanto2023flexible}
N.~Accanto, F.~G. Blot, A.~Lorca-C{\'a}mara, V.~Zampini, F.~Bui, C.~Tourain,
  N.~Badt, O.~Katz, and V.~Emiliani, ``A flexible two-photon fiberscope for
  fast activity imaging and precise optogenetic photostimulation of neurons in
  freely moving mice,'' {\em Neuron}, vol.~111, no.~2, pp.~176--189, 2023.

\bibitem{lucas1981iterative}
B.~D. Lucas and T.~Kanade, ``An iterative image registration technique with an
  application to stereo vision,'' in {\em IJCAI'81: 7th international joint
  conference on Artificial intelligence}, vol.~2, pp.~674--679, 1981.

\bibitem{hansen1987truncated}
P.~C. Hansen, ``The truncated svd as a method for regularization,'' {\em BIT
  Numerical Mathematics}, vol.~27, pp.~534--553, 1987.

\bibitem{malone2023diffuserspec}
J.~D. Malone, N.~Aggarwal, L.~Waller, and A.~K. Bowden, ``Diffuserspec:
  spectroscopy with scotch tape,'' {\em Optics Letters}, vol.~48, no.~2,
  pp.~323--326, 2023.

\bibitem{weber1995robust}
J.~Weber and J.~Malik, ``Robust computation of optical flow in a multi-scale
  differential framework,'' {\em International Journal of Computer Vision},
  vol.~14, no.~1, pp.~67--81, 1995.

\bibitem{papadopoulos2016apollon}
D.~Papadopoulos, J.~Zou, C.~Le~Blanc, G.~Ch{\'e}riaux, P.~Georges, F.~Druon,
  G.~Mennerat, P.~Ramirez, L.~Martin, A.~Fr{\'e}neaux, {\em et~al.}, ``The
  apollon 10 pw laser: experimental and theoretical investigation of the
  temporal characteristics,'' {\em High Power Laser Science and Engineering},
  vol.~4, p.~e34, 2016.

\bibitem{ranc2022etude}
L.~Ranc, {\em Etude et am{\'e}lioration spatio-temporelles de l'intensit{\'e}
  et du contraste dans les lasers PW}.
\newblock PhD thesis, universit{\'e} Paris-Saclay, 2022.

\bibitem{sai2020designing}
T.~Sai, M.~Saba, E.~R. Dufresne, U.~Steiner, and B.~D. Wilts, ``Designing
  refractive index fluids using the kramers--kronig relations,'' {\em Faraday
  discussions}, vol.~223, pp.~136--144, 2020.

\bibitem{mounaix2020time}
M.~Mounaix, N.~K. Fontaine, D.~T. Neilson, R.~Ryf, H.~Chen, J.~C.
  Alvarado-Zacarias, and J.~Carpenter, ``Time reversed optical waves by
  arbitrary vector spatiotemporal field generation,'' {\em Nature
  communications}, vol.~11, no.~1, p.~5813, 2020.

\bibitem{chen2022synthesizing}
L.~Chen, W.~Zhu, P.~Huo, J.~Song, H.~J. Lezec, T.~Xu, and A.~Agrawal,
  ``Synthesizing ultrafast optical pulses with arbitrary spatiotemporal
  control,'' {\em Science Advances}, vol.~8, no.~43, p.~eabq8314, 2022.

\bibitem{cruz2022synthesis}
D.~Cruz-Delgado, S.~Yerolatsitis, N.~K. Fontaine, D.~N. Christodoulides,
  R.~Amezcua-Correa, and M.~A. Bandres, ``Synthesis of ultrafast wavepackets
  with tailored spatiotemporal properties,'' {\em Nature Photonics}, vol.~16,
  no.~10, pp.~686--691, 2022.

\bibitem{Papadopoulos_2022}
J.~P. Zou, H.~Coïc, and D.~Papadopoulos, ``Spatiotemporal coupling
  investigations for ti:sapphire-based multi-pw lasers,'' {\em High Power Laser
  Science and Engineering}, vol.~10, p.~e5, 2022.

\bibitem{sung2023hyperspectral}
Y.~Sung, ``Hyperspectral three-dimensional refractive-index imaging using
  snapshot optical tomography,'' {\em Physical Review Applied}, vol.~19, no.~1,
  p.~014064, 2023.

\bibitem{curry2011direct}
N.~Curry, P.~Bondareff, M.~Leclercq, N.~F. Van~Hulst, R.~Sapienza, S.~Gigan,
  and S.~Gr{\'e}sillon, ``Direct determination of diffusion properties of
  random media from speckle contrast,'' {\em Optics letters}, vol.~36, no.~17,
  pp.~3332--3334, 2011.

\bibitem{shinn2023phantom}
M.~Shinn, ``Phantom oscillations in principal component analysis,'' {\em
  Proceedings of the National Academy of Sciences}, vol.~120, no.~48,
  p.~e2311420120, 2023.

\bibitem{DataBase}
``Refractive index database.'' \url{https://refractiveindex.info}.

\end{thebibliography}


\clearpage

\newpage

\onecolumn

\appendix

\beginsupplement

\bigskip
\huge {A Polarimetric Wavefront Imager: Supplementary Materials }\\[1em]
\large{Baptiste Blochet,$^{1,\ast}$ Nathalie Lebas,$^{2}$ Pascal Berto,$^{1,3,4}$ Dimitrios Papadopoulos,$^{2}$ Marc Guillon$^{1,4,5,+}$}\\[1em]
\noindent
\normalsize{$^{1}$ Saints-Pères Paris Institute for the Neurosciences, CNRS UMR 8003, Universit{\' e} Paris Cit{\' e},}\noindent\\
\normalsize{ 45 rue des Saints-Pères, Paris 75006, France}\\
\normalsize{$^{2}$LULI-CNRS, CEA, Sorbonne Universite, Ecole Polytechnique, Institut Polytechnique de Paris, Palaiseau 91128, France}\\
\normalsize{$^{3}$Institut de la Vision, Sorbonne Université, CNRS-UMR 7210, Inserm-UMR S968, Paris 75012, France}\\
\normalsize{$^{4}$Institut Universitaire de France, Paris, France}\\
\normalsize{$^{5}$Institut Langevin, ESPCI Paris, Universit\'e PSL, CNRS, Paris 75005, France}\\
\begin{tabular}{rl}
\normalsize{E-mails:} & \normalsize{$^\ast$ baptiste.blochet@u-paris.fr}\\
  & \normalsize{$^+$ marc.guillon@u-paris.fr}\\
\end{tabular}
\\[3em]


\section{Table of specifications of mulispectral and hyperspectral WFS implementations}

\begin{table*}[h!]
\centering
\begin{tabular}{ |p{5cm}||p{4cm}|p{4cm}|p{4cm}|  }
 \hline
  & Metrology&Apollon&Microscopy\\
 \hline
 d-maps calibration   & Average over entire field of view    & Average over entire field of view & estimated at every camera pixels\\
 \hline
Preprocessing &   & Patch based cross correlation & \\
 \hline
Matrix inversion & Truncated SVD $\sigma_{SVD }= 30$ & Truncated SVD $\sigma_{SVD }= 30$ & Moore-Penrose pseudo inverse \\
 \hline
 Multiscale filter width vector $\sigma_{scale}$ & [6,3,0] pixels & [0] pixels & [0,0,0,0,0] pixels \\
 \hline
 Patch size & Square patches of 128x128 pixels & Square patches of 128x128 pixels & Sliding Gaussian window ($\sigma = 4~{\rm pixels}$)\\ 
 \hline
 FOV size at the MCF plane & $0.8\times0.8~{\rm mm^2}$ & $0.8\times0.8~{\rm mm^2}$ & $1.2\times1.2~{\rm mm^2}$ \\
 \hline
 FOV size at the sample plane & $2.4\times2.4~{\rm mm^2}$ & $160\times160~{\rm mm^2}$ & $120\times120~{\rm \mu m^2}$ \\
 \hline
 Spectral resolution & $5~{\rm nm}$& $5~{\rm nm}$& 3 channels \\
 \hline
 Spatial resolution at the MCF plane & $66~{\rm \mu m}$ & $66~{\rm \mu m}$ & $4.5~{\rm \mu m}$ \\
 \hline
 Distance imaged MCF - Camera  & $88-97~{\rm mm}$ & $~88-97~{\rm mm}$ & $~4-8~{\rm mm}$ \\
 \hline
 Sensitivity & $0.35~{\rm \mu rad}$ & $0.35~{\rm \mu rad}$ & $7.5~{\rm \mu rad}$ \\
 \hline
\end{tabular}
 \caption{\textbf{Summary of the relevant reconstruction and physical parameters}}
 \label{table:algo}
\end{table*}

\twocolumn

\section{Distance calibration maps} 

\begin{figure}[h!]
\centering
\includegraphics[width=1\linewidth]{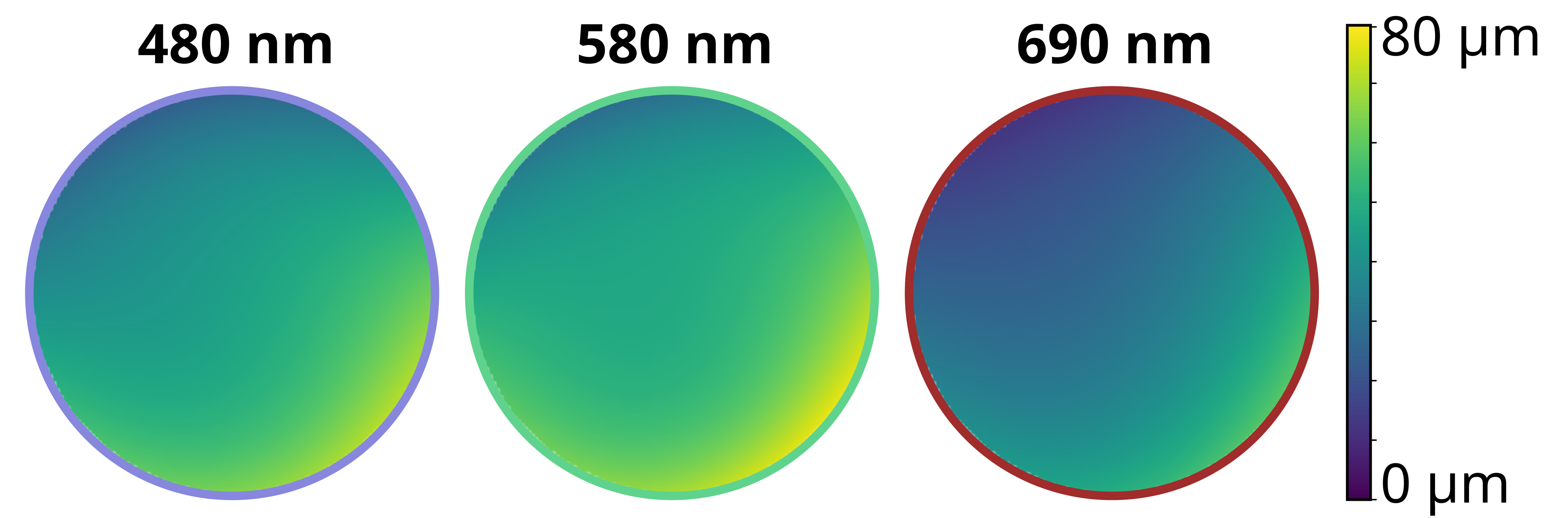}d
\caption{\textbf{Distance $d$ between the output facet of the MCF and the imaged plane on the camera (in $\mu m$) including the field curvature at the three considered wavelengths.}  
}
\label{fig:zmaps}
\end{figure}

For the multispectral wavefront microscopy, the distance between the bundle output and the image plane on camera was experimentally measured at each of the three wavelength by scanning a point source, located at the focal point of a fixed lens, with a translation stage. For each position of the point source, yielding a tilted collimated beam, the displacements were measured at the camera.
At each camera pixel, the displacement curve was fitted with a affine function as a function of the input angle. The distance $d$ was estimated according to Eq.~\ref{eq:grad_displacement}. The distance maps $d(x,y,\lambda)$ was then fitted by two-dimensional polynomials of order 3 for each wavelength $\lambda$. The results d calibration maps are shown in Fig.~\ref{fig:zmaps}. 

\section{Sample and refractive index model}

Poly(methyl methacrylate) beads of $5.2~{\rm \mu m}$ (PMMA-R-5.2, microparticles GmbH) were deposited on glass microscope slides. The dry beads were then covered with a solution consisting in blue brilliant absorber FCF in water (E133 in water, Vahine, Avignon, France) mixed with sucrose ($2~{\rm g/ml}$) and then sealed with a cover glass. We model the refractive index of the solution as~\cite{sai2020designing}
\begin{equation}
n_{solution} = n_{water} + n_{sucrose} + A.n_{FCF}
\end{equation}
where $n_{sucrose}$ is assumed to be a non-dispersive constant index offset over the entire spectrum. $n_{FCF}$ was computed based on the normalized absorption spectrum according to the Kramer-Kronig relations. The factor coefficient $A$, proportional to the (unknown) FCF concentration, was fitted based on refractive indices measured with Abbe refractometer.
The refractive indices of the beads ($n_{PMMA}$) and water ($n_{water}$) were calculated from Sellmeier models \cite{DataBase} 


\section{Supplementary figures}

 \begin{figure*}[h!]
    \centering
    \includegraphics[width=1\linewidth]{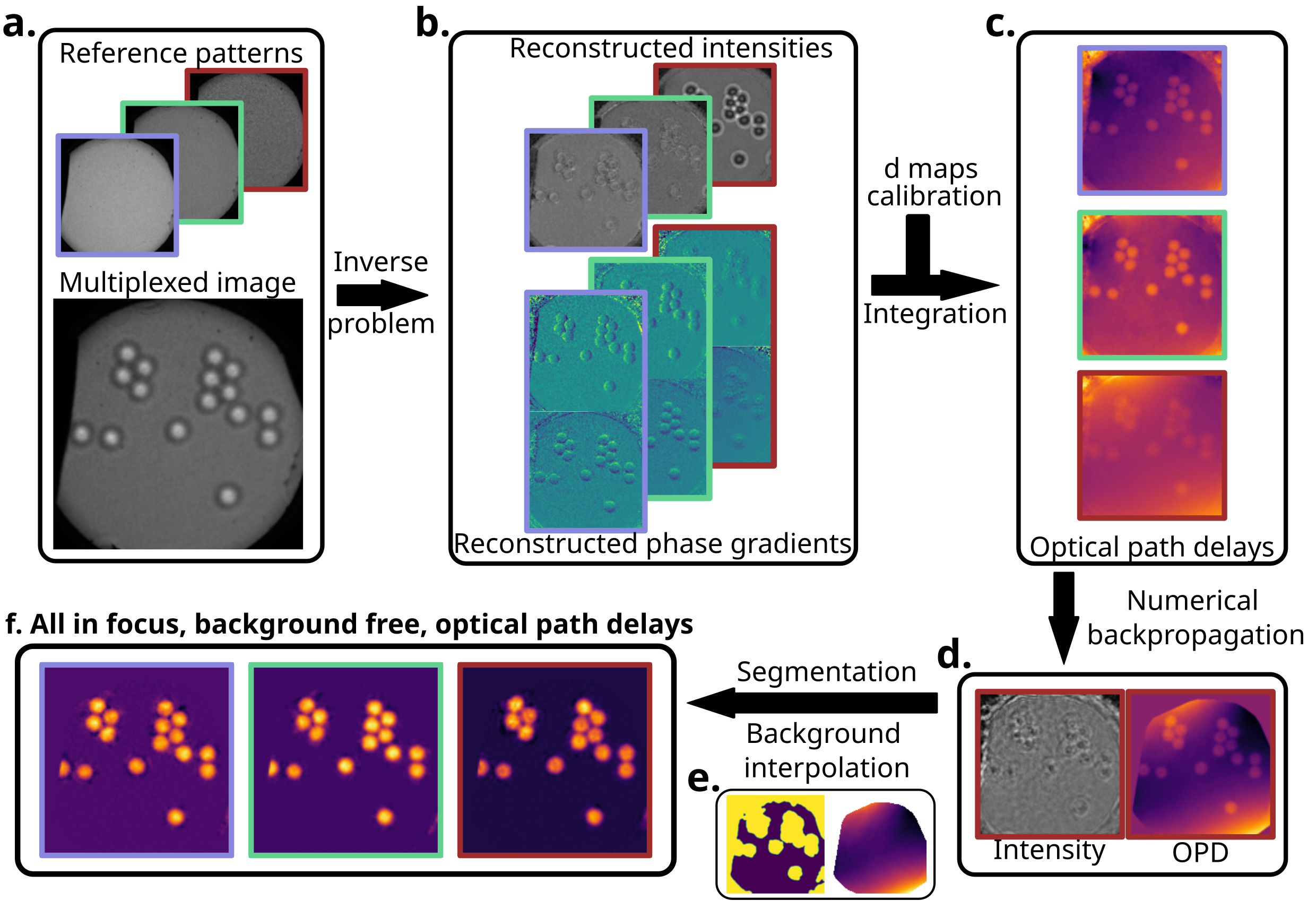}
    \caption{\textbf{Post processing of the multispectral microscopy images . a. } Initial references images measured under plane wave illumination at 480, 580 and 690 nm and multiplexed image acquired under broad illumination. \textbf{b.} Reconstructed intensities and phase gradients computed by inverted Eq.~(4). \textbf{c.} Optical paths delays generated after numerical integration according to Eq.~(3). \textbf{d.} Numerical back propagation were achieved for the three wavelengths by minimizing the variance of the intensities images with Fresnel propagation. \textbf{e.} The low varying optical path delays backgrounds were estimated at the position of the PMMA beads by interpolation after segmentation. \textbf{f.} All in focus, background free optical path delays images were obtained after background subtraction.  
}
    \label{fig:postproc}
\end{figure*}

\begin{figure*}[h!]
\centering
\includegraphics[width=1\linewidth]{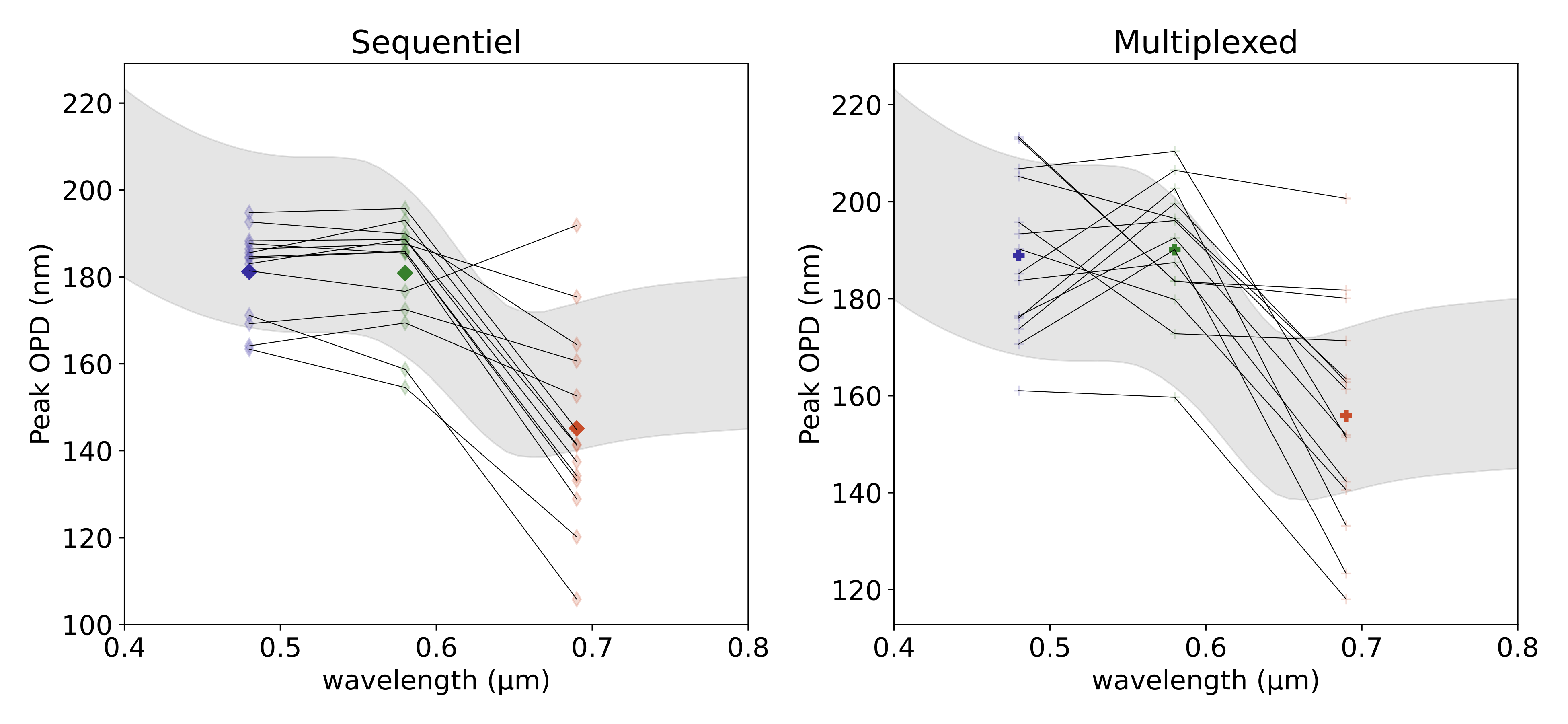}
\caption{\textbf{Comparison of peaks OPDs measured from sequential and multiplexed reconstructions }  
}
\label{fig:peak_OPD}
\end{figure*}

\pagebreak
\newpage

\end{document}